# Common 0.1 bar Tropopause in Thick Atmospheres Set by Pressure-Dependent Infrared Transparency

T. D. Robinson[1,2]*, D. C. Catling[2,3,4]

**A minimum atmospheric temperature, or tropopause, occurs at a pressure of around 0.1 bar in the atmospheres of Earth[1], Titan[2], Jupiter[3], Saturn[4], Uranus and Neptune[4], despite great differences in atmospheric composition, gravity, internal heat and sunlight. In all these bodies, the tropopause separates a stratosphere with a temperature profile that is controlled by the absorption of shortwave solar radiation, from a region below characterised by convection, weather, and clouds[5,6]. However, it is not obvious why the tropopause occurs at the specific pressure near 0.1 bar. Here we use a physically-based model[7] to demonstrate that, at atmospheric pressures lower than 0.1 bar, transparency to thermal radiation allows shortwave heating to dominate, creating a stratosphere. At higher pressures, atmospheres become opaque to thermal radiation, causing temperatures to increase with depth and convection to ensue. A common dependence of infrared opacity on pressure, arising from the shared physics of molecular absorption, sets the 0.1 bar tropopause. We hypothesize that a tropopause at a pressure of approximately 0.1 bar is characteristic of many thick atmospheres, including exoplanets and exomoons in our galaxy and beyond. Judicious use of this rule could help constrain the atmospheric structure, and thus the surface environments and habitability, of exoplanets.**

"Atmospheric structure" is usually taken to mean an average vertical temperature profile, which provides fundamental information about how physical and chemical processes change with altitude. In discussing atmospheric structure, the terms *tropopause* and *radiative-convective boundary* are sometimes used interchangeably. Here, we keep these terms separate because they are located at different levels in the real atmospheres that we consider, with Titan being the most extreme case. Specifically, we use 'tropopause' to mean the temperature minimum. Remote sensing and *in situ* measurements have shown that tropopauses all occur around ~0.1 bar on planets in the Solar System with thick atmospheres and stratospheric inversions (Fig. 1). No explanation exists for this common tropopause level, so we investigated its physics with an analytic one-dimensional model of atmospheric structure described in ref. 7. Dynamics can modulate the tropopause pressure with latitude[8,9], but radiative-convective equilibrium exerts first-order control of globally averaged structure.

[1]NASA Ames Research Center, Moffett Field, California 94035, USA, [2]NASA Astrobiology Institute's Virtual Planetary Laboratory, University of Washington, Seattle, Washington 98195, USA, [3]University of Washington Astrobiology Program, University of Washington, Seattle, Washington 98195, USA, [4]Department of Earth and Space Sciences, University of Washington, Seattle, Washington 98195, USA.
*email: tyler.d.robinson@nasa.gov

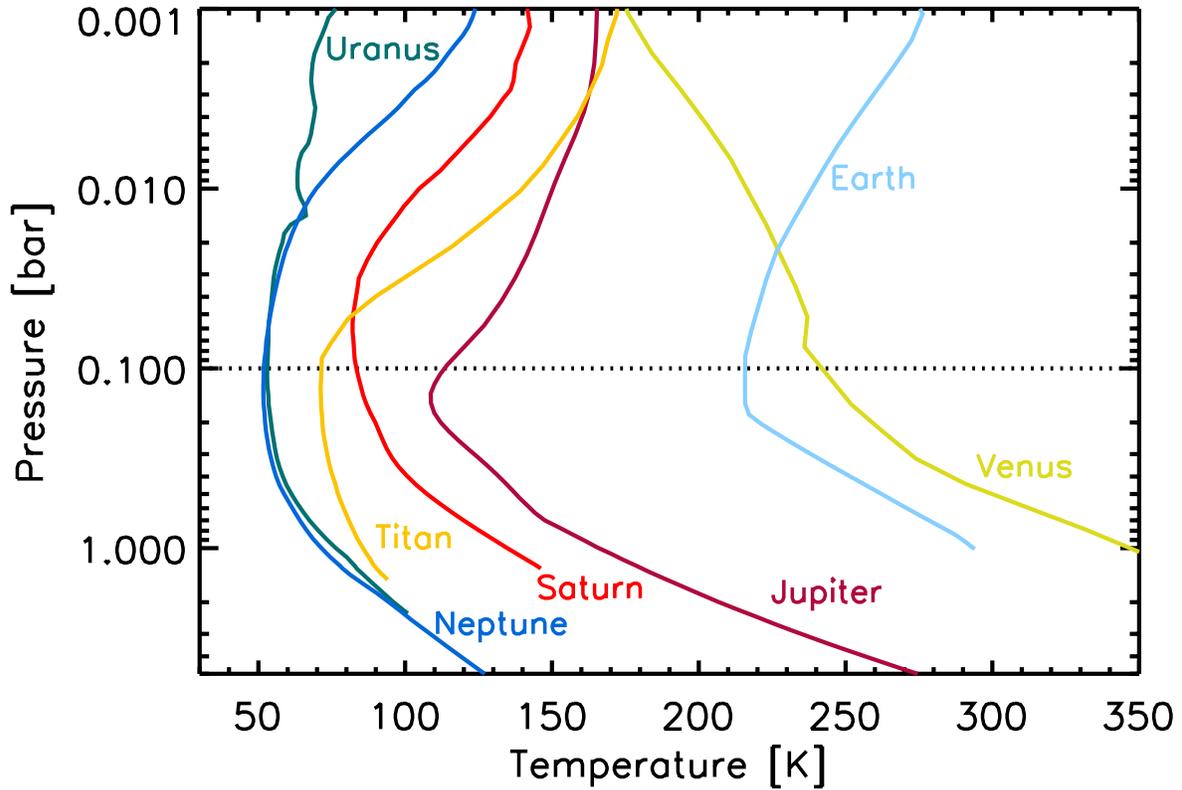

**Figure 1 | Temperature-pressure profiles for worlds in the Solar System with thick atmospheres[1-4,28]**. *Temperature minima commonly occur around 0.1 bar. Venus has a very weak 0.1 bar tropopause in the global mean (see text). More information regarding data sources is given in the Supplementary Materials.*

Our model is constructed as follows. Infrared opacities are grey, i.e., described by a single, broadband optical depth, $\tau_{IR}$, at every pressure level. Solar radiative transfer occurs in stratospheric and tropospheric channels. Parameters $k_{strato} = \tau_{sws}/\tau_{IR}$ and $k_{tropo} = \tau_{swt}/\tau_{IR}$ control the attenuation of solar energy in these two channels, where $\tau_{sws}$ and $\tau_{swt}$ are shortwave optical depths. Tropospheric convection follows a dry adiabat adjusted by an empirical scaling factor typically around 0.6-0.9 to match an observed mean moist adiabat in each atmosphere. The ratio ($\gamma$) of specific heats at constant pressure ($c_p$) and volume ($c_v$), respectively, ($\gamma = c_p/c_v$) sets the dry adiabatic lapse rate, and is 1.4 for atmospheres dominated by diatomic gases, like those considered here.

Our model uses a known power law between pressure ($p$) and the grey infrared optical depth of $\tau_{IR} \propto p^n$ (ref. 10). This scaling arises from combining the differential optical depth $d\tau_{IR} = -\kappa \rho_a dz$ (where $\kappa$ is a grey opacity, $\rho_a$ is the absorber mass density and $dz$ is the differential altitude) with hydrostatic equilibrium $dp/dz = -g\rho$ (where $g$ is the gravitational acceleration and $\rho$ is atmospheric density), so that $d\tau_{IR} \propto \kappa dp$ (see Supplementary Information). Below middle stratospheres, radiative transfer is dominated by pressure-broadening and collision-induced absorption, which have $\kappa \propto p$, and, thus, $n = 2$ from integration[11]. We do not use $n = 1$, which would correspond to higher levels of the atmosphere where Doppler broadening dominates[12,13] and $\kappa$ is independent of pressure (see also Fig. S3).

Given $\gamma$, $n$, and other inputs (Table 1), our model computes radiative-convective equilibrium by solving for infrared optical depths at the radiative-convective boundary ($\tau_{rc}$) and a reference pressure ($\tau_0$, determined at either the surface or 1 bar where the atmosphere is optically thick in the infrared). The greenhouse effect necessary to maintain temperature $T_0$ at $p_0$ is related to $\tau_0$. Consequently, the tropopause pressure, $p_{tp}$, weakly declines with increasing greenhouse effect according to $p_{tp} \propto \tau_0^{-1/n} \propto \tau_0^{-1/2}$, consistent with studies of Earth's contemporary warming[14].

**Table 1 | Parameters and computed variables for simple models of atmospheric structure of Earth, Jupiter, Saturn, Titan, Uranus, and Neptune.** *Parameters include a tropospheric reference pressure ($p_0$) and temperature ($T_0$), an adjustment to the dry adiabat to account for volatile condensation ($\alpha$), the internal energy flux ($F_i$), the solar flux absorbed in the stratosphere ($F^\odot_{strato}$) and troposphere ($F^\odot_{tropo}$), and two parameters that control solar energy attenuation in these two channels ($k_{strato}$ and $k_{tropo}$). For rocky bodies, $F^\odot_{tropo}$ includes surface absorption. For example, Earth has $F^\odot_{strato} + F^\odot_{tropo} = 240$ W/m², the global mean absorbed solar flux. See the Supplementary Materials for more information.*

| World | Earth | Jupiter | Saturn | Titan | Uranus | Neptune |
|---|---|---|---|---|---|---|
| $p_0$ [bar] | 1 | 1 | 1 | 1.5 | 1 | 1 |
| $T_0$ [K] | 288 | 166 | 135 | 94 | 76 | 72 |
| $\alpha$ | 0.6 | 0.85 | 0.94 | 0.77 | 0.83 | 0.87 |
| $F^\odot_{strato}$ [W/m²] | 7 | 1.3 | 0.41 | 1.49 | 0.24 | 0.09 |
| $F^\odot_{tropo}$ [W/m²] | 233 | 7.0 | 2.04 | 1.12 | 0.41 | 0.18 |
| $F_i$ [W/m²] | negligible | 5.4 | 2.01 | negligible | negligible | 0.43 |
| $k_{strato}$ | 90 | 90 | 180 | 120 | 220 | 580 |
| $k_{tropo}$ | 0.16 | 0.06 | 0.03 | 0.20 | 0.08 | 0.20 |
| $\tau_{rc}$ | 0.15 | 0.34 | 0.44 | 4.4 | 0.62 | 0.41 |
| $\tau_0$ | 1.9 | 6.3 | 9.2 | 5.6* | 8.7 | 3.0 |
| $\tau_{tp}$ | 0.050 | 0.064 | 0.040 | 0.077 | 0.042 | 0.017 |
| $p_{tp}$ [bar] (model) | 0.16 | 0.10 | 0.066 | 0.18 | 0.070 | 0.075 |
| $p_{tp}$ [bar] (observed) | 0.16 | 0.14 | 0.08 | 0.2 | 0.11 | 0.1 |

*For Titan, $\tau_0$=2.5 at $p$=1 bar.

Modeling results have common features shown schematically in Fig. 2, with precise values given in Table 1. First, tropopause pressures are all correctly computed near 0.1 bar. Second, the radiative-convective boundary is always below the tropopause. The scaled optical depth at the radiative-convective boundary is $D\tau_{rc} \approx 1$ for all worlds except Titan (the diffusivity factor, $D$, is ~1.66 and accounts for the integration of radiance over a hemisphere[15,16]). Titan has a very shallow convective region up to only ~1.3 bar because significant shortwave absorption in its upper hazy troposphere causes stability against convection. Third, the grey infrared optical depth at 1 bar tends to range between 2 and 10 (mean = 5.3 ±3.2). Finally, the grey infrared optical depth of the tropopause is always about $D\tau_{tp} \approx 0.1$ (mean = 0.08±0.03).

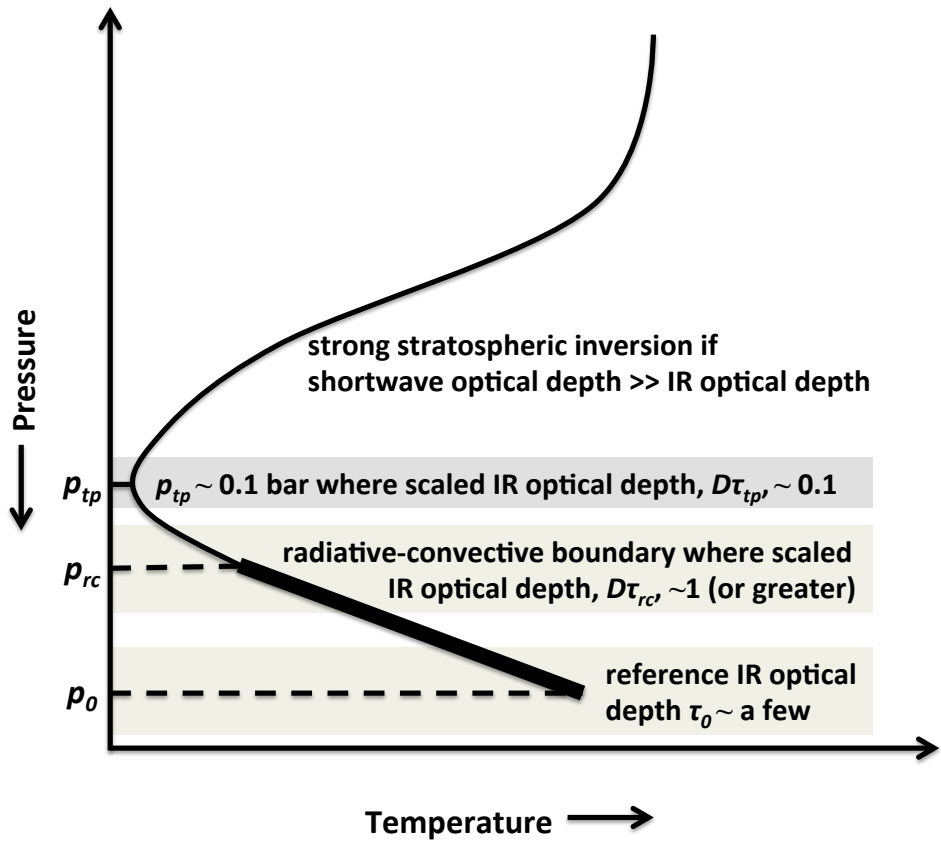

**Figure 2 | Schematic diagram of thermal structure in a thick planetary atmosphere with a stratospheric inversion.** *A general feature is a grey infrared optical depth, $\tau_0$, of order a few at a pressure, $p_0$, of 1 bar. There is a radiative-convective boundary at a scaled infrared optical depth, $D\tau_{rc}$, of about unity or greater. A tropopause temperature minimum occurs at a pressure, $p_{tp}$, of about 0.1 bar and a scaled infrared optical depth, $D\tau_{tp}$, of about 0.1. The "diffusivity factor" D for the optical depth is ~1.66 (see text). The thickened portion of the profile indicates the convective part of the troposphere.*

The tropopause temperature minimum occurs in the radiative regime, where an analytic expression for the temperature (equation S2) allows us to assess the conditions for the minimum. Setting the derivative of the temperature profile to zero gives the infrared optical depth at the tropopause as:

$$\tau_{tp} = \frac{1}{k_{strato}} \ln\left[\frac{F^\odot_{strato}}{F^\odot_{tropo} + F_i}\left(\frac{k^2_{strato}}{D^2} - 1\right)\right]. \quad (1)$$

Here, $F^\odot_{strato}$ and $F^\odot_{tropo}$ are the solar fluxes absorbed in the stratosphere and troposphere, respectively, while $F_i$ is the internal heat flux of the planet. (For planets with surfaces, $F^\odot_{tropo}$ is the solar flux absorbed at the surface and in the troposphere). Note that equation (1) does not depend on the ratio of specific heats ($\gamma$). Consequently, the inferred tropopause level is valid for thick atmospheres with tropopause minima dominated by either triatomic or diatomic gases (e.g., $CO_2$ or $H_2$, respectively). The expression only depends on two parameters: the ratio $F^\odot_{strato}/(F^\odot_{tropo} + F_i)$ and $k_{strato}$. The former gives the ratio of the stratospheric absorbed flux relative to that from below, which in Table 1 is highest for Titan and lowest for Earth. Essentially, $k_{strato}$ parameterizes the effect of an arbitrary shortwave absorber through an exponential decline in heating from a stratopause.

Equation (1) yields a non-zero, physical result only when

$$k^2_{strato} > D^2\left[1 + (F^\odot_{tropo} + F_i)/F^\odot_{strato}\right], \quad (1)$$

which sets the threshold value of $k_{strato}$ for the formation of a tropopause temperature minimum and stratospheric inversion. Values of $k_{strato}$ above this threshold cause deposition of enough energy at low pressures for a stratospheric inversion. If $k_{strato}$ is below the threshold, shortwave energy is absorbed at depths where infrared radiation primarily determines the temperature structure, thus preventing the formation of an inversion and tropopause minimum (see Supplementary Information). Typical flux ratios in this expression mean that $k_{strato}$ must be of order ~100 for a well-developed minimum. Also the larger the internal flux, $F_i$, the larger $k_{strato}$ needs to be for an inversion to exist.

Contours in a plot of stratospheric attenuation $k_{strato}$ versus $F^\odot_{strato}/(F^\odot_{tropo} + F_i)$ show the range of $D\tau_{tp}$ for the bodies of the Solar System (Fig. 3a). The plot covers a broader range of parameter space than these bodies alone and demonstrates a general rule that worlds with relatively strong stratospheric inversions will tend to have $D\tau_{tp} \approx 0.1$ (see Supplementary Information Fig. S1).

Venus does not have a well-developed tropopause temperature minimum in the global average because it lacks a significant stratospheric inversion, which is consistent with our tropopause theory. However, Venus is marginal (Fig. 3a) and in fact, possesses a distinct tropopause temperature minimum at ~0.1 bar in its mid to high latitudes[17], and so conforms to the ~0.1 bar rule when a minimum is seen (see Supplementary Information). The reason for the latitudinal variations in tropopause sharpness is unknown but may be a modulation of the radiative-convective mean state by a Hadley-like meridional circulation above the cloud tops[18]. The interpretation is complicated by the presence of unknown absorbers at 0.2-0.5 μm (ref. 19). Mars' low surface pressure of ~0.006 bar means that it does not fall within our scope of examining commonalities in thick atmospheres and so Mars is not plotted in Fig. 3a. However, the lack of a shortwave absorber (i.e., $k_{strato}$ << 1) accounts for Mars' absence of a stratospheric inversion.

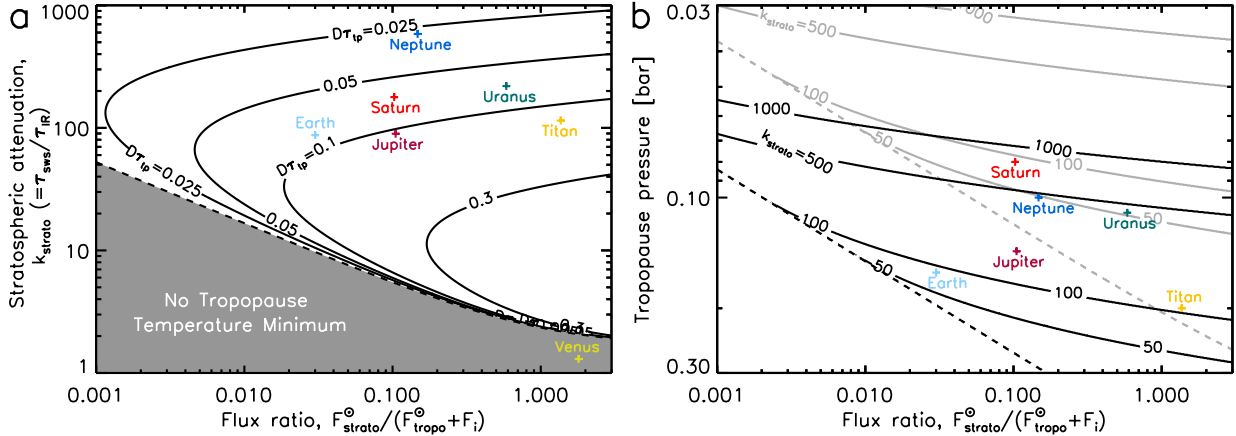

**Figure 3 | Tropopause grey infrared optical depth and pressure. a,** *Contours of $D\tau_{tp}$ over a wide range of parameter space for the stratospheric attenuation strength parameter ($k_{strato}$) and the ratio of the stratospheric absorbed stellar flux ($F^{\odot}_{strato}$) to the sum of the tropospheric absorbed stellar flux and the internal heat flux ($F^{\odot}_{tropo} + F_i$).* **b,** *Tropopause pressures (from equations (1) and (3)) over a wide range of values for the ratio of $F^{\odot}_{strato}$ to ($F^{\odot}_{tropo} + F_i$). Contours are for $k_{strato}$, where black lines assume $\tau_0$=2 at 1 bar, while grey lines assume $\tau_0$=10 at 1 bar. Values for Solar System worlds are indicated.*

While optical depths near the 1 bar pressure levels generally lie between 2 and 10 for the bodies in the Solar System, line-by-line calculations justify the same range for a suite of Titan-like $N_2$-$CH_4$-$H_2$ atmospheres encompassing surface temperatures of 80-140 K, $H_2$-dominated gas giant atmospheres over a range of 1 bar reference temperatures of 75-400 K, and Earth-like $N_2$-$H_2O$-$CO_2$ atmospheres with surface temperatures from 250-300 K (see Supplementary Information). The Earth-like atmospheres are of particular interest for exoplanets because a "habitable" planet is conventionally defined as one where liquid water is stable on the surface[20].

We can now see how ~0.1 bar tropopauses arise for worlds with atmospheric compositions like those of the Solar System. Because pressure-broadening or collision-induced absorption apply generally to thick atmospheres, we can use the scaling between pressure and grey infrared optical depth to relate the tropopause pressure, $p_{tp}$, to $\tau_{tp}$ via

$$p_{tp} = p_0 \left( \tau_{tp} / \tau_0 \right)^{0.5} . \qquad (2)$$

Using our constraint that $D\tau_{tp} \approx 0.1$ and the range of values for $\tau_0$ at a reference pressure of 1 bar ($2 \leq \tau_0 \leq 10$), the tropopause pressure must be near 0.1 bar (e.g., the average atmosphere of Table 1 has $p_{tp} \approx (0.05/5)^{0.5}$). Figure 3b uses the bounding values of $\tau_0$ (2 and 10) at $p_0$ = 1 bar to show tropopause pressures near 0.1 bar for parameter space that encompasses and goes beyond the Solar System bodies.

As a consequence of the above calculations, we hypothesize that a ~0.1 bar tropopause is an emergent rule that will apply to thick atmospheres on numerous exoplanets and exomoons that have compositions that are not markedly dissimilar to those in the Solar System, like the oxidizing conditions of Earth and Venus, or the reducing conditions of Titan and the giant planets. For the latter, carbon will be in the form of $CH_4$, which along with photochemically-generated hydrocarbons will create an inversion[21]. Amongst oxidizing stratospheres, an $O_2$-$CO_2$ atmosphere has ozone, while a $CO_2$-$SO_3$ Venus-like stratosphere will have elemental sulfur shortwave absorbers (ref. 21, p. 291) (see Table S1).

Our hypothesis—which clearly only applies to atmospheres with stratospheric inversions—is testable. Already, observations of HD 209458b, a so-called "Hot Jupiter" that is thought to possess a stratospheric inversion, yield rough estimates for a tropopause minimum near 0.1 bar[22]. Although the rule works for Titan, which has a 16 day rotational period, it may prove inappropriate to apply globally averaged models to rotationally locked bodies, which could possess strong temperature contrasts between day and night hemispheres. Similarly, a runaway greenhouse with large infrared opacity will not satisfy equation (1) for a stratospheric inversion, so is outside the scope of our rule.

Exceptions aside, our proposed rule could help with the assessment of future telescopic spectra of exoplanet atmospheres[23,24] in many ways. For example, an algorithm for retrieving atmospheric properties has recently been applied to synthetic spectra of an Earth-like world[25]. In many cases, the retrieved tropopause pressure was near 1 bar, which was much deeper than the 0.3 bar "target" tropopause used to generate spectra. Such algorithms could be improved by assuming a 0.1 bar tropopause as a Bayesian prior.

Because of its applicability to Earth-like worlds, the 0.1 bar tropopause rule could help in assessing exoplanet habitability. In the future, surface pressure could be estimated from fits to exoplanet spectral features[26] while the surface temperature might be obscured because of the lack of an infrared window at the wavelengths of observation. Alternatively, the surface temperature might be estimated from a spectrum while the surface pressure remains unknown. In either scenario, a 0.1 bar tropopause assumption in a radiative-convective model[7] would allow an estimate of surface temperature or pressure, respectively, which together are required to assess liquid water stability. The tropopause temperature would be needed. This could come from spectral features but, if not, a reasonable first order estimate is the "skin temperature" of $T_{eff}/2^{0.25}$, where $T_{eff}$ is the effective blackbody temperature of the planet (ref. 6, p.404). For Earth, for example, $T_{eff}$ = 255 K and the skin temperature is 214 K, which is within a few percent of the observed global mean tropopause at ~208 K (ref. 27).

Thus, a unity of physics not only explains ~0.1 bar tropopauses in thick Solar System atmospheres but also has the implication of potentially constraining exoplanet habitability.

**Acknowledgements**

This work was performed as part of the NASA Astrobiology Institute's Virtual Planetary Laboratory, supported by the National Aeronautics and Space Administration through the NASA Astrobiology Institute under solicitation No. NNH05ZDA001C.  TDR gratefully acknowledges support from an appointment to the NASA Postdoctoral Program at NASA Ames Research Center, administered by Oak Ridge Associated Universities.  DCC was also supported by NASA Exobiology/Astrobiology grant NNX10AQ90G.  The authors thank the late Conway Leovy for discussions in which he was supportive of pursuing the idea that a 0.1 bar tropopause constitutes an emergent law.

**Author contributions**

TDR and DCC made equally important contributions to the project and co-wrote the paper.  TDR generated the model outputs.

**Additional information**

Supplementary information is available in the online version of this paper.  Reprints and permissions information is available online at www.nature.com/reprints.  Correspondence and requests for materials should be addressed to TDR.

**Competing financial interests**

The authors declare no competing financial interests.

**References**
1	McClatchey, R. A., Fenn, R. W., Selby, J. E. A., Volz, F. E. & Garing, J. S. Optical Properties of the Atmosphere (Third Edition). (Air Force Cambridge Research Labs, 1972).
2	Lindal, G. F. *et al.* The atmosphere of Titan - an analysis of the Voyager 1 radio occultation measurements. *Icarus* **53**, 348-363, doi:10.1016/0019-1035(83)90155-0 (1983).
3	Moses, J. I. *et al.* Photochemistry and diffusion in Jupiter's stratosphere: Constraints from ISO observations and comparisons with other giant planets. *Journal of Geophysical Research (Planets)* **110**, E08001, doi:10.1029/2005JE002411 (2005).
4	Lindal, G. F. The atmosphere of Neptune - an analysis of radio occultation data acquired with Voyager 2. *Astron J* **103**, 967-982, doi:10.1086/116119 (1992).
5	Sanchez-Lavega, A. *An Introduction to Planetary Atmospheres*.  (CRC Press/Taylor & Francis, 2010).
6	Goody, R. M. & Yung, Y. L. *Atmospheric Radiation: Theoretical Basis*.  (Oxford University Press, 1989).
7	Robinson, T. D. & Catling, D. C. An Analytic Radiative-Convective Model for Planetary Atmospheres. *The Astrophysical Journal* **757**, 104 (2012).
8	Schneider, T. The tropopause and the thermal stratification in the extratropics of a dry atmosphere. *J Atmos Sci* **61**, 1317-1340 (2004).

9       Haqq-Misra, J., Lee, S. & Frierson, D. M. W. Tropopause structure and the role of eddies. *J Atmos Sci* **68**, 2930-2944 (2011).
10      Pollack, J. B. Temperature structure of nongray planetary atmospheres. *Icarus* **10**, 301-313 (1969).
11      McKay, C. P., Lorenz, R. D. & Lunine, J. I. Analytic Solutions for the Antigreenhouse Effect: Titan and the Early Earth. *Icarus* **137**, 56-61, doi:10.1006/icar.1998.6039 (1999).
12      Kondrat'ev, K. I. A. *Radiation in the Atmosphere*. Vol. 12 (Academic Press, 1969).
13      Hanel, R. A., Conrath, B. J., Jennings, D. E. & Samuelson, R. E. *Exploration of the Solar System by Infrared Remote Sensing*. (Cambridge University Press, 2003).
14      Seidel, D. J. & Randel, W. J. Variability and trends in the global tropopause estimated from radiosonde data. *J Geophys Res* **111**, D21101 (2006).
15      Armstrong, B. H. Theory of the diffusivity factor for atmospheric radiation. *Journal of Quantitative Spectroscopy and Radiative Transfer* **8**, 1577-1599, doi:10.1016/0022-4073(68)90052-6 (1968).
16      Rodgers, C. D. & Walshaw, C. D. Polynomial Approximations to Radiative Functions. *Quarterly Journal of the Royal Meteorological Society* **89**, 422-423 (1963).
17      Tellmann, S., Patzold, M., Hausler, B., Bird, M. K. & Tyler, G. L. Structure of the Venus neutral atmosphere as observed by the Radio Science experiment VeRa on Venus Express. *J. Geophys. Res.* **114**, E00B36 (2009).
18      Newman, M., Schubert, G., Kliore, A. J. & Patel, I. R. Zonal winds in the middle atmosphere of Venus from Pioneer Venus radio occultation data. *Journal of Atmospheric Sciences* **41**, 1901-1913 (1984).
19      Mills, F. P., Esposito, L. W. & Yung, Y. L. in *Exploring Venus as a Terrestrial Planet Geophysical Monograph Series* (eds L.W. Esposito, E. R. Stofan, & T. E. Cravens)  73-100 (American Geophysical Union, 2007).
20      Kasting, J. F., Whitmire, D. P. & Reynolds, R. T. Habitable zones around main sequence stars. *Icarus* **101**, 108-128 (1993).
21      Yung, Y. L. & DeMore, W. B. *Photochemistry of Planetary Atmospheres*. (Oxford Univ. Press, 1999).
22      Madhusudhan, N. & Seager, S. A Temperature and Abundance Retrieval Method for Exoplanet Atmospheres. *Astrophysical Journal* **707**, 24-39, doi:10.1088/0004-637X/707/1/24 (2009).
23      Deming, D. *et al.* Discovery and Characterization of Transiting Super Earths Using an All-Sky Transit Survey and Follow-up by the James Webb Space Telescope. *Publications of the Astronomical Society of the Pacific* **121**, 952-967, doi:10.1086/605913 (2009).
24      Beichman, C. A., Woolf, N. J. & Lindensmith, C. A. *The Terrestrial Planet Finder (TPF): A NASA Origins Program to Search for Habitable Planets*. (NASA Jet Propulsion Laboratory, 1999).
25      von Paris, P., Hedelt, P., Selsis, F., Schreier, F. & Trautmann, T. Characterization of potentially habitable planets: Retrieval of atmospheric and planetary properties from emission spectra. *Astronomy & Astrophysics* **551**, A120 (2013).
26      Des Marais, D. J. *et al.* Remote Sensing of Planetary Properties and Biosignatures on Extrasolar Terrestrial Planets. *Astrobiology* **2**, 153-181 (2002).

27    Han, T. T., Ping, J. S. & Zhang, S. J. Global features and trends of the tropopause derived from GPS/CHAMP RO data. *Science China: Physics, Mechanics & Astronomy* **54**, 365-374 (2011).
28    Moroz, V. I. & Zasova, L. V. VIRA-2: a review of inputs for updating the Venus International Reference Atmosphere. *Advances in Space Research* **19**, 1191-1201, doi:10.1016/S0273-1177(97)00270-6 (1997).